\documentclass[floatfix,amssymb,prd,twocolumn,superscriptaddress,nofootinbib, aps]{revtex4-1}
\usepackage{lineno}
\usepackage{bbding}
\usepackage{pifont}
\usepackage{amsmath}
\usepackage{amssymb}
\usepackage{mathtools}
\usepackage{tikz}
\usetikzlibrary{calc}
\usepackage{tikz-feynman}\tikzfeynmanset{compat=1.1.0}
\usepackage{slashed}
\usepackage{bbm}
\usepackage{dsfont}
\usepackage{xcolor}
\usepackage[normalem]{ulem}
\usepackage{graphicx}
\usepackage{subcaption}
\usepackage{orcidlink}
\usepackage{braket}

\tikzset{
  solid/.style={line width=0.7pt},
  dashed/.style={line width=0.7pt, densely dashed},
}

\newcommand{\Tr}{\text{Tr}}

\newcommand{\hf}{\frac{1}{2}}
\newcommand{\lag}{\mathcal{L}}
\newcommand{\tx}[1]{\text{#1}}

\newcommand{\ep}{\epsilon}
\usepackage{tikz}
\usetikzlibrary{calc}

\newcommand{\tilx}{\tilde{X}}

\begin{document}

\title{\boldmath Hidden Zeros in Massive Theories}

\author{Mariana Carrillo Gonzalez\orcidlink{0000-0003-1119-9097}}
\email{m.carrillo-gonzalez@imperial.ac.uk \\ m.carrillo-gonzalez@soton.ac.uk}
\affiliation{Abdus Salam Centre for Theoretical Physics, Imperial College, London, SW7 2AZ, U.K.}
\affiliation{School of Physics \& Astronomy, University of Southampton,
Highfield, Southampton, SO17 1BJ, U.K.}
\author{Freddie Ward}
\email{freddie.ward23@imperial.ac.uk}
\affiliation{Abdus Salam Centre for Theoretical Physics, Imperial College, London, SW7 2AZ, U.K.}

\begin{abstract}
We investigate whether the hidden zeros and associated factorisations found for massless colour-ordered amplitudes persist under massive deformations. Using the kinematic mesh construction, we show that hidden zeros survive only for symmetry controlled mass generation. For massive $\Tr \Phi^3$ with a uniform mass, the zeros and their factorisation patterns are inherited after a massive shift of planar variables, and an analogous statement holds for Kaluza-Klein reductions where the relevant non-planar variables are modified by conserved mode numbers. For the non-linear sigma model (NLSM), a naive pion mass term generically spoils hidden zeros, while a spurion induced potential restores them. This allows factorisation near zeros, including odd point channels described by an appropriately mass deformed NLSM + $\phi^3$ theory, and leads to a hidden zero based on-shell recursion for massive NLSM amplitudes. For spin-one, a simple massive Yang-Mills theory fails to exhibit hidden zeros, while spontaneously broken gauge theories preserve them.
\end{abstract}

\maketitle
\flushbottom

\section{Introduction}

Understanding the consequences of the physical properties of scattering amplitudes, such as unitarity, causality, and improved ultraviolet behaviour, can lead to useful insights and tools like BCFW recursion relations, generalized unitarity, and positivity bounds \cite{Elvang:2015rqabook, deRham:2022hpx, Dixon:1996wi, Travaglini:2022uwo, Mizera:2023tfe, ZhiboedovNotesAnalyticSmatrix2023, Bern:2011qt}. Recently, some novel factorizations of scattering amplitudes were discovered away from physical poles \cite{Cachazo:2021wsz,HZ,Cao:2024gln} which has led to a quest to understand their physical origin and the theories which have these properties. 

In this letter, we will focus on the construction of~\cite{HZ}, where it was discovered that $\Tr \Phi^3$, the non-linear sigma model (NLSM), and Yang-Mills-scalar\footnote{This theory corresponds to gluons coupled to complex, massless adjoint scalars with a quartic interaction and arises from a dimensional reduction of YM \cite{Arkani-Hamed:2023swr}} theory (YMS) all share a set of \emph{hidden zeros}: points in the space of Lorentz-invariant dot products of momenta at which partial amplitudes vanish. Near these zeros, the amplitudes factorise into three pieces involving known scattering amplitudes. This behaviour was observed in earlier explorations of dual resonant amplitudes in \cite{DAdda:1971wcy} and also arises for uncoloured theories obtainable via the double copy \cite{zerosfromdoublecopy,Li:2024qfp,Cao:2024gln}. It has now been extended to a wide range of theories beyond those considered initially, and its origin and implications have been extensively explored \cite{Guevara:2024nxd, Chang:2025cqe,Rodina_2025, Cao:2024qpp,Zhou:2024ddy,Zhang:2024efe,Li:2024bwq,Huang:2025blb, Zhou:2025tvq, Zhang:2025zjx,De:2025bmf,Feng:2025ofq,Jones:2025rbv,Li:2025suo,Backus:2025hpn}. 

The goal of this letter is to determine whether hidden zeros and factorisations survive under massive deformations. We will show below that this holds only for specific massive deformations in which symmetries are broken in a controlled way. Both spontaneous symmetry breaking and the breaking of symmetries through spurions give rise to theories in which hidden zeros and factorisations survive, with newly defined planar and non-planar kinematic variables.

\paragraph*{Conventions} We work with mostly minus signature and with the generators of the $U(N)$ and $SU(N)$ groups satisfying $\Tr(T^aT^b)=\delta^{ab}$ and $
[T^{a}, T^{b}] = i \sqrt{2}\, f^{abc}\, T^{c}$.

\section{Hidden Zeros and the Kinematic Mesh} \label{sec:hz_massless}
The tree level amplitudes of coloured theories can be decomposed as a sum over partial amplitudes $A_n [\sigma_1 \dots \sigma_n]$ with single-trace coefficients. For the scattering of $n$ particles with momenta $k_i$, $i = 1,\dots, n$, partial amplitudes of any scalar theory for the canonical ordering $[12,\dots n]$ can be written in terms of  the $n(n-3)/2$ independent planar variables\footnote{Here we assume that we work in $d$ dimensions with $d>n$ to avoid additional constraints arising from Gram determinants.}
\begin{equation}
    X_{i,j} \coloneq (p_i + \dots + p_{j-1})^2 \ ,
\end{equation}
where subscripts are $\mod{n}$. For massless theories with local interactions, these planar variables are simply the denominators of propagators found in planar channels. By momentum conservation $\sum_i p_i = 0$ and the on-shell condition for massless particles, $p_{i}^{2} = 0$, the planar variables satisfy $X_{i, i+1} = X_{1,n} = 0$. The planar variables are related to a set of non-planar variables  $c_{i,j} \coloneq - 2 p_i \cdot p_j$,
with $i,j$ non-adjacent, through the $c$-equation:
\begin{equation}\label{cequation}
    c_{i,j} = X_{i,j} + X_{i+1,j+1} - X_{i+1,j} - X_{i,j+1} \ .
\end{equation}
These non-planar variables are, in the massless case, simply equivalent to the Mandelstam variables since $s_{ij} \coloneq (p_i + p_j)^2=- c_{i,j}$.

\begin{figure}[t!]
    \centering

    \begin{subfigure}[t]{\columnwidth}
        \centering
        \includegraphics[width=0.6\columnwidth,trim={0 6pt 0 6pt},clip]{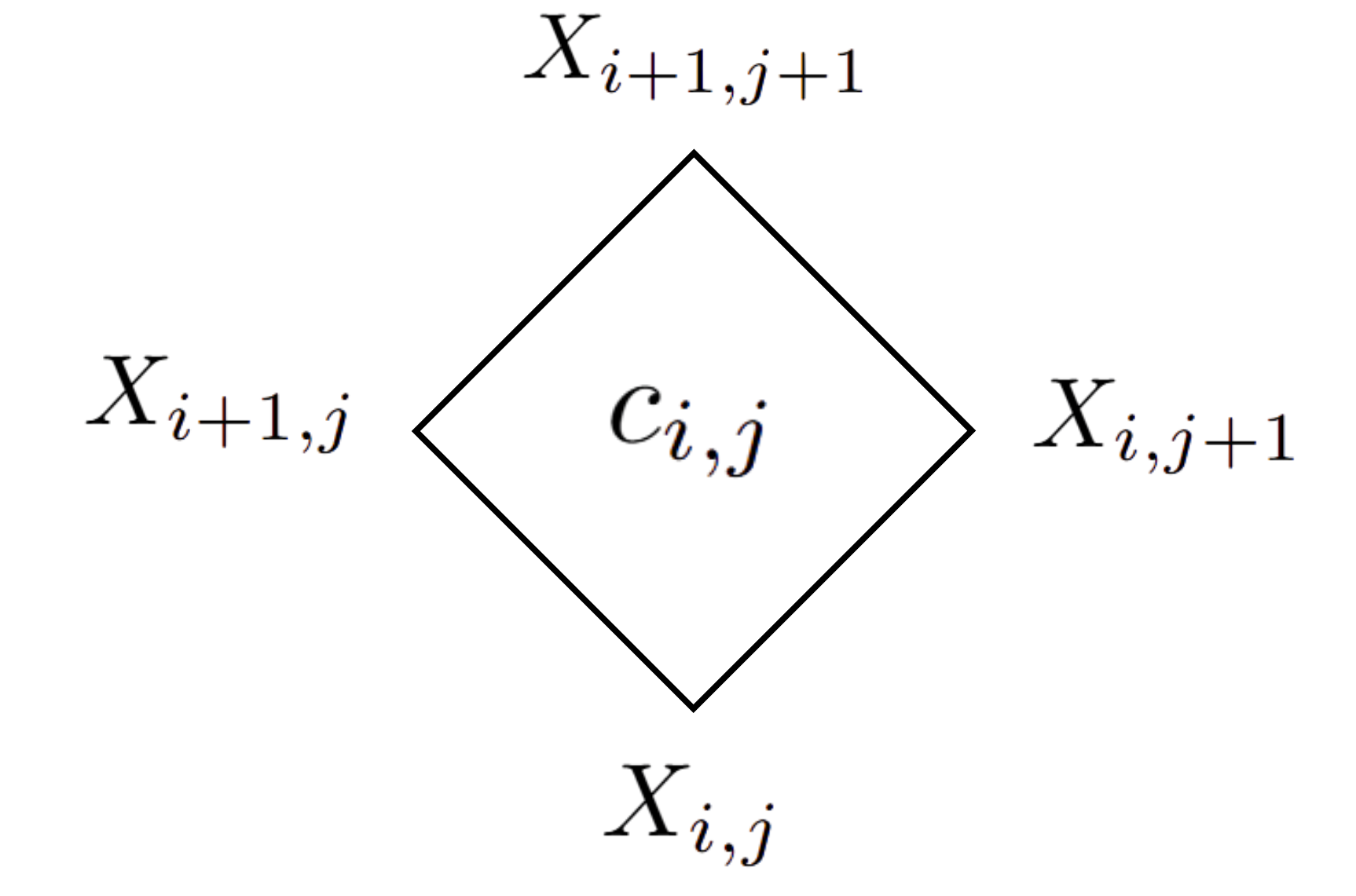}
        \label{fig:mesh-lef}
    \end{subfigure}

    \begin{subfigure}[t]{\columnwidth}
        \centering
        \includegraphics[width=0.78\columnwidth,trim={0 6pt 0 6pt},clip]{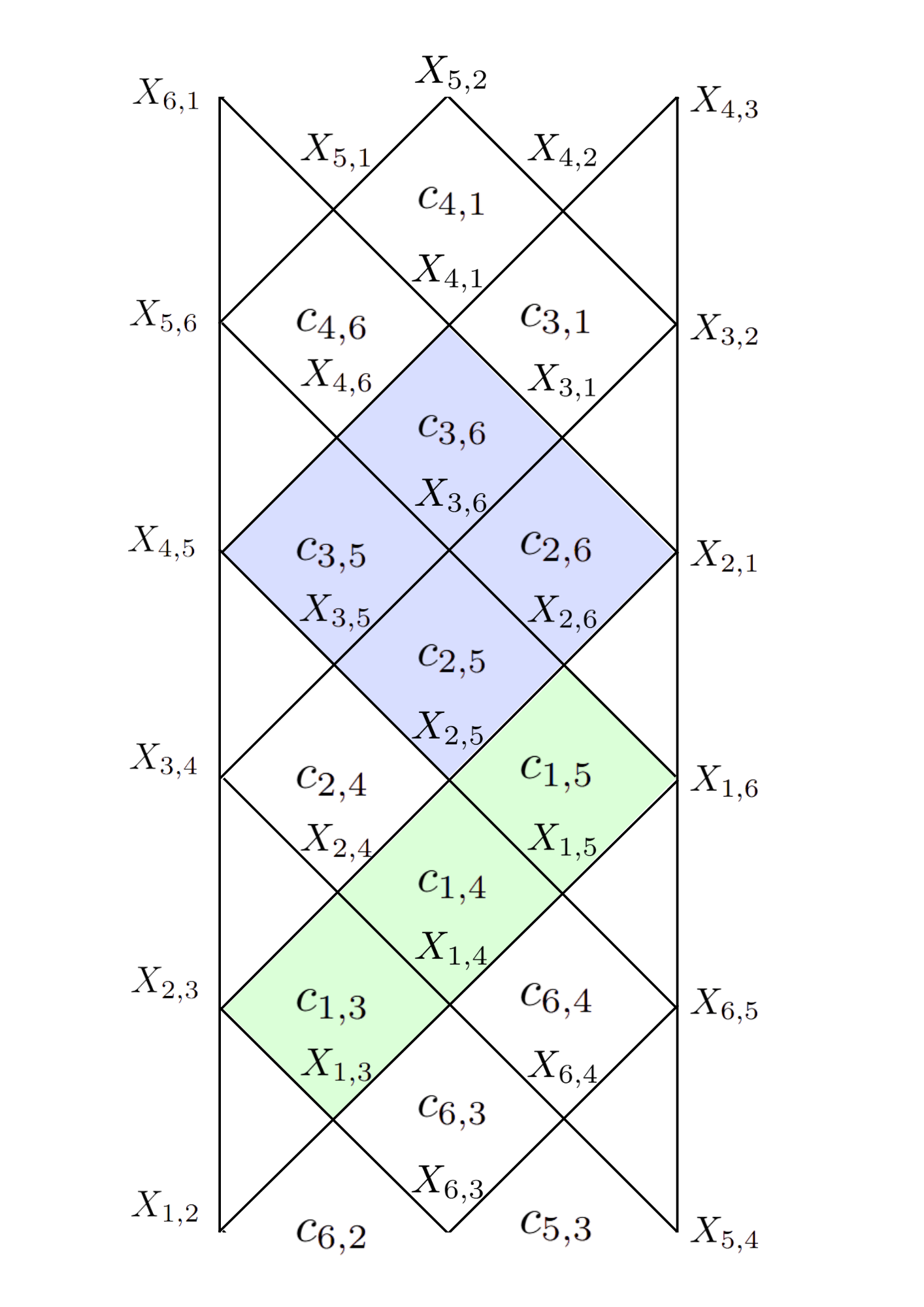}
        \label{fig:mesh-right}
    \end{subfigure}

    \vspace{-4pt}
    \caption{\textbf{Top}: Tiles that make up the kinematic mesh. \textbf{Bottom}: Examples of causal diamonds on the $n=6$ kinematic mesh. Setting all $c_{i,j}=0$ inside the causal diamonds makes the corresponding 6-point amplitude vanish.}
    \vspace{-6pt}
    \label{fig:mesh-main}
\end{figure}

Planar and non-planar variables can be neatly organised using the \emph{kinematic mesh} \cite{HZ, arkanihamed2022causaldiamondsclusterpolytopes}. To construct the mesh, we associate to each non-planar $c_{i,j}$ a diamond, the four vertices of which correspond to the four planar variables in Eq.(\ref{cequation}), as in the top of Fig.~\ref{fig:mesh-main}. The diamonds then join together to tile a region with grid points corresponding to planar variables and tiles to non-planar variables, with boundaries corresponding to the vanishing planar variables. An example of a mesh for $n=6$ scattering is shown in the bottom of Fig.~\ref{fig:mesh-main}. Taking any rectangle drawn on the mesh, we have that 
\begin{equation}
X_T + X_B - X_L - X_R \;=\; \sum_{(ij)\in \mathrm{interior}} c_{i,j} \ ,
\end{equation}
where on the left-hand side we have the planar variables associated with the four vertices of the rectangle, and on the right-hand side, the sum is over all non-planar variables contained within its interior. 
Hidden zeros lie at points where a subset of the non-planar variables are set to zero and are found especially efficiently using the kinematic mesh: for scalar amplitudes, setting all the $c_{i,j}$ to zero inside a maximal rectangle, with $X_L$ and $X_R$ on the boundaries, called a causal diamond, sets the corresponding amplitude to zero; examples are shown in Fig.~\ref{fig:mesh-main}. These zeros also have a geometric origin understood through the ABHY associahedron~\cite{ABHY}: they correspond to the limit in which the polytope collapses to one of a lower dimension~\cite{HZ}. 

For an $n$-point scalar amplitude, the set of hidden zeros can be translated from the set of causal diamonds on the mesh as \cite{Backus:2025hpn}
\begin{align} \label{eq:hz}
c_{i,j} &= 0 \quad  \text{where} \nonumber \\
&i \in \{ m, m+1, \dots, m+k-1 \}, \nonumber \\
&j \in \{ k+m+1, k+m+2, \dots, m-2+n \},
\end{align}
with $m = 1 , \dots , n$ and $k = 1, \dots , n-4$. This way, we can classify each zero by a pair of numbers $(k,m)$, where $k$ corresponds to the number of tiles on the bottom-left edge of the associated causal diamond and $m$ is the index shared by all the $c_{i,j}$ on the diamond's lowest-right row. Many of the zeros are equivalent or cyclically equivalent. For amplitudes with external gauge fields, there is the important caveat that a hidden zero must also come with similar statements about the polarisation vectors. Specifically, a hidden zero in YM takes the form 
\begin{equation}\label{spin1zeroconditions}
c_{i,j} = \ep_i \cdot \ep_j = \ep_i \cdot p_j = \ep_j \cdot p_i = 0 \ ,
\end{equation}
where $i,j$ are given as in Eq.~\eqref{eq:hz}. 

An interesting aspect of hidden zeros is the property that near a zero amplitudes will factorise in a predictable way. If we take a particular zero but turn back on one of the non-planar variables, denoted $c_{*}$, the amplitude factorises as
\begin{equation}\label{generalfactorisation}
    A_{n}[12 \dots n] \xrightarrow{c_{*} \neq 0} \left( \frac{c_{*}}{X_{T} X_{B}} \right) \times A^{\tx{up}} \times A^{\tx{down}},
\end{equation}
where $X_{T}$ and $X_{B}$ are the top and bottom vertices of the causal diamond in the mesh that defines the zero, and the ``up" and ``down" sub-amplitudes are tree-level amplitudes whose explicit form depends on the original zero, the chosen $c_{*}$, and the theory under consideration. 

\begin{figure}[t!]
\includegraphics[width=0.8\columnwidth]{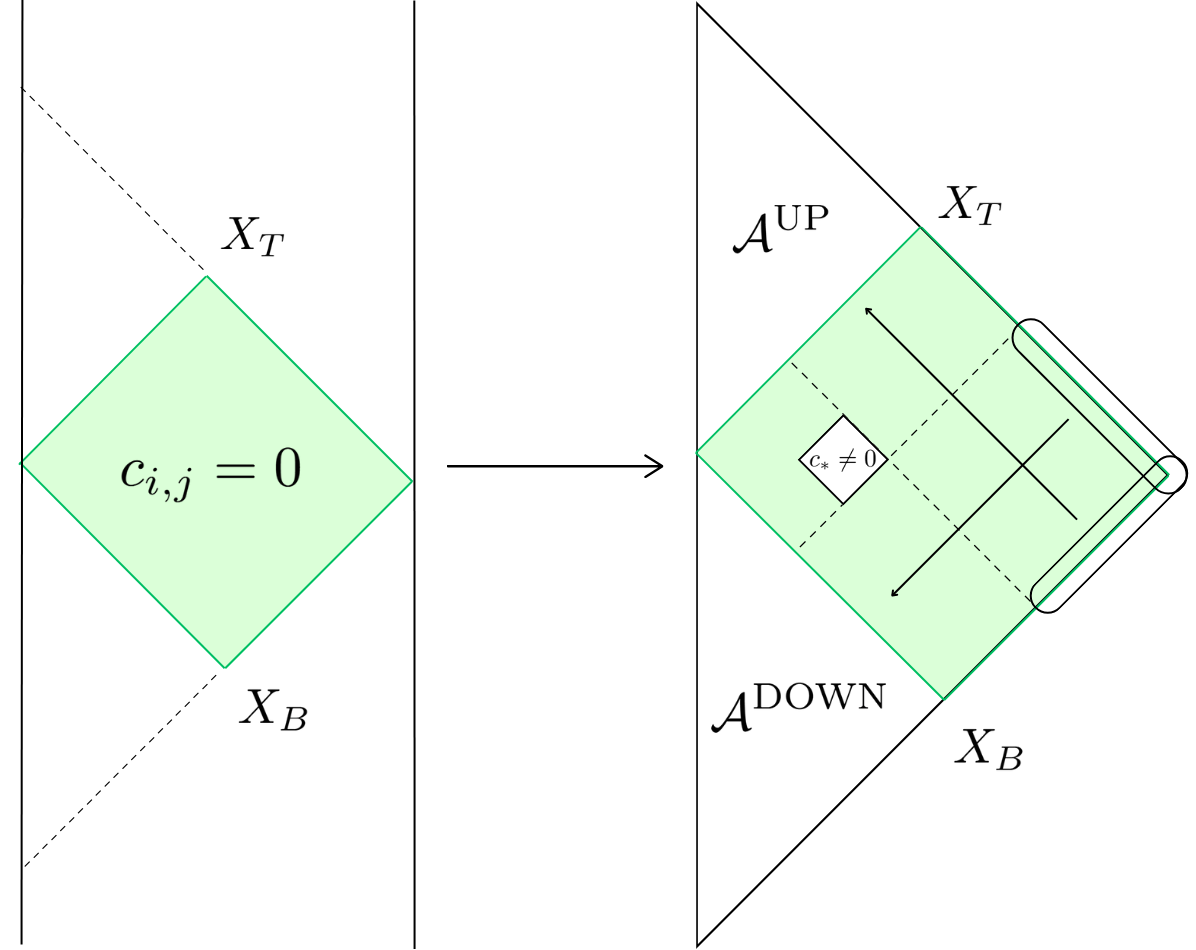}
    \caption{General scheme for constructing the factorisation in Eq.~(\ref{generalfactorisation}). The two triangular regions formed by extending the causal diamond to the mesh boundary define regions with variables defining lower-point processes. The planar variables in the green lower and upper edges of the ``up" and ``down" regions respectively are replaced by planar variables opposite to them in the encircled edge.}
    \label{fig: Factorisation Phi3}
\end{figure}

For $\Tr \Phi^3$ the sub-amplitudes are again tree-level $\Tr \Phi^3$ amplitudes and the factorisation pattern falls into one of two categories \cite{Jones:2025rbv}: associating an index pair $i,j$ to a causal diamond, and hence a zero, through $X_{B} = X_{i,j}$ we have either (1) that $c_{*}$ is the right-most tile in the causal diamond, in which case the up and down amplitudes take the form 
\begin{equation}
    \begin{aligned}
        &A^{\tx{up}} \quad = A[j-1, j, \dots, i-2, i-1] \\
        &A^{\tx{down}} = A[i, i+1, \dots , j-1, j],
    \end{aligned}
\end{equation}
or (2) $c_{*}$ is any other tile, i.e., $c_{*} = c_{k,l}$ with $i \leq k \leq j-2$ and $j \leq l \leq i-2$, in which case the up and down sub-amplitudes have a subset of their planar variables exchanged for others in the following way: 
\begin{equation}
    \begin{aligned}
        A^{\tx{up}} &= A[j-1, j, \dots, i-2, i-1]\Big|_{X_{j-1, b} \rightarrow X_{i,b}}  \\
        A^{\tx{down}} &= A[i, i+1, \dots , j-1, j] \Big|_{X_{a,j} \rightarrow X_{a, i-1}},
    \end{aligned}
\end{equation}
where $a=i+1, \dots, k$ and $b = l+1, \dots, i-2$. This variable exchange can be read off from the kinematic mesh, as in Fig.~\ref{fig: Factorisation Phi3}. The story is the same for the NLSM if both the up and down amplitudes are even-point; the amplitude factorises as per Eq.~(\ref{generalfactorisation}) with NLSM sub-amplitudes. But if the factorisation pattern results in odd-point factors then these will be amplitudes of a mixed theory of pions and cubic bi-adjoint scalars \cite{Cachazo_2016_cubicandnlsm}, and the pattern changes to
\begin{equation}\label{odd-point NLSM factorisation}
    A_n \xrightarrow{c_{*} \neq 0} \left( X_{T} + X_{B} \right) \times A^{\tx{up}, \tx{NLSM} + \phi^3} \times A^{\tx{down}, \tx{NLSM} + \phi^3},
\end{equation}
where the up and down amplitudes are still determined through the mesh, but with additional rules to determine the external configuration of the cubic scalars and pions as described in \cite{HZ}. In both cases, we still exchange planar variables depending on the particular $c_{*} \neq 0$ as in $\Tr \Phi^3$. When the hidden zero condition is relaxed with one $c_{*} \neq 0$ we have that $X_{T} + X_{B} = c_{*}$, which again makes the zero manifest in Eq.~(\ref{odd-point NLSM factorisation}) in the limit $c_{*} \rightarrow 0$. 

%-----------------------------------
\section{Massive \texorpdfstring{$\Tr \Phi^3$}{phi3} Theory}\label{Section: Massive phi^3 zeros}

We now generalise the previous properties to massive theories, starting with massive $\Tr \Phi^3$. Our first example will be a simple deformation where all fields acquire a mass $m$. To analyse this, we first need to adjust the variables used to write our amplitudes. Planar variables $X_{i,j}$ were a natural choice of basis over kinematic space in the massless theory since they appear in the denominators of propagators and exhibit cyclicity. In keeping with this logic, we define massive planar variables intuitively as massive planar propagator denominators as
\begin{equation}\label{Massive Planar Variable Definition}
    \tilde{X}_{i,j} \coloneq X_{i,j} - m^2.
\end{equation}
Momenta are now massive with $p_i^2 = m^2$ and so we have boundary values analogous to the massless case: $ \tilx_{i,i+1} = \tilx_{1,n} = 0$. Motivated by the kinematic mesh construction, we define tilded non-planar variables in analogy to Eq.~(\ref{cequation}) as
\begin{equation}\label{tilde X c equation}
    \tilde{c}_{i,j}  \coloneq \tilx_{i,j} + \tilx_{i+1, j+1} - \tilx_{i+1, j} - \tilx_{i, j+1} \ ,
\end{equation}
which in this case are equivalent to the massless ones
\begin{equation}
   \tilde{c}_{i,j} = c_{i,j}
\end{equation}
By virtue of this, it is simple to see that the massive kinematic mesh is constructed in an identical manner to the massless theory. Any massive amplitude, when written in terms of massive planar variables $\tilx_{i,j}$, will be identical in \emph{form} to the massless amplitudes since there are no derivative interactions. This, coupled with the fact that the relation between the non-planar and planar variables for the massive case remains unchanged, means that all zeros and factorisations from the massless theory carry through to the massive theory.  For example, in the simplest four-point case we have that 
\begin{equation}
    A_{4}^{\tx{massive}} [1234] = \frac{1}{\tilx_{1,3}} + \frac{1}{\tilx_{2,4}} = \frac{c_{1,3}}{\tilx_{1,3} \tilx_{2,4}} \xrightarrow{c_{1,3} = 0} 0.
\end{equation}
Using the definition of the non-planar variables, we can see that in the massive case they no longer correspond to negative Mandesltams, but are now related to them as $s_{ij} = - c_{i,j} +2m^2$. This means that a non-planar being set to zero does not imply the corresponding Mandelstam is zero. For instance, for the above four-point example the zero locus corresponds to $s_{13} = 2m^2$.

Meanwhile, the factorisation generalizes straightforwardly to
\begin{equation}\label{massive factorisation general}
\begin{aligned}
     A^{\tx{massive}}_{n} &\xrightarrow{c_{*} \neq 0} \left( \frac{1}{X_{T} - m^2} + \frac{1}{X_{B} - m^2} \right) \times A^{\text{up}} \times A^{\text{down}}, \\
     &= \left( \frac{c_{*}}{\tilx_{T} \tilx_{B}} \right) \times A^{\text{up}} \times A^{\text{down}},
     \end{aligned}
\end{equation}
where $A^{\text{up}/\text{down}}$ are massive amplitudes associated to the massive kinematic mesh constructed out of the $\tilx_{i,j}$. 

We can also ascribe an ABHY associahedron to the massive theory. The associahedron for the massive theory is the same as the massless associahedron but its relation to the planar variables $X$ is translated in each direction by $m^2$ \cite{Jagadale:2022rbl,WardMScThesis2025HiddenZerosMassive}, so all statements that lead to factorisation still apply, they just occur at a positions in kinematic space uniformly shifted. Even more transparently, the massive associahedron construction written in terms of $\tilx_{i,j}$ variables is identical to the massless associahedron written in terms of $X_{i,j}$ variables. 

%-----------------------------------
\section{Dimensional Reduction of \texorpdfstring{$\Tr \Phi^3$}{Tr phi3}}\label{Section: KK phi^3}

In this section, we analyse another mass deformation of $\Tr \Phi^3$ given by the the Kaluza-Klein (KK) reduction of this theory. We start with the massless theory in $d+1$-dimensions and compactify the $d+1$ dimension around a circle so that the background spacetime has the product structure $\mathds{R}^{d-1,1} \times S^{1}$. 

In $d+1$ dimensions, the massless action is 
\begin{equation}
    S(x^{\mu} , y) = \int d^{d+1}x \, \Tr(\partial^{M} \Phi \partial_{M} \Phi) + g \Tr ( \Phi^3 ),
\end{equation}
where $y \sim y + 2 \pi R$ is the coordinate over $S^{1}$ with $R$ its radius, and $x^{M} = (x^{\mu}, y)$. As usual, we expand the components of the scalar field in terms of Fourier modes as $\Phi^{a}(x^M)
 = \sum_{n\in\mathds{Z}} e^{iny/R}\,\Phi_{n}^{a}(x^\mu)$, where $n \in \mathds{Z}$, and find that the $d$-dimensional action is
\begin{align}
  &S =\! \int \!\!d^{d}x \sum_{n}\!\!
  \left(\!
    \partial_{\mu}\Phi_{n}^{a}(x)\,
    \partial^{\mu}\Phi_{-n}^{a}(x)
    - \left(\tfrac{n}{R}\right)^{2}
      \Phi_{n}^{a}(x)\,\Phi_{-n}^{a}(x)\!
  \right)
  \nonumber \\
  &
  + \tilde{g}\, C^{abc} \!\!\!\sum_{n_{1},n_{2},n_{3}}\!\!\!
    \delta_{n_{1}+n_{2}+n_{3},0} \;
    \Phi_{n_{1}}^{a}(x)\,
    \Phi_{n_{2}}^{b}(x)\,
    \Phi_{n_{3}}^{c}(x),
\end{align}
where $C^{abc}=\Tr(T^aT^bT^c)$ and we have redefined the fields to absorb a constant from the $S^{1}$ integration. This action contains an infinite tower of massive modes with masses $m_n^2 = m^2 n^2$, where $m = 1/R$, and a cubic interaction which imposes a constraint on the mode numbers of the three fields at each vertex. The conservation of the integer mode number reflects a global $U(1)$ symmetry associated with translations along the compact $S^1$ direction. Given this, the fields appearing in propagators have mediating masses $m^2_{i,j} \coloneq m^2 (n_i + \dots + n_{j-1})^2$. 

To analyse the zeros, we define a new set of planar variables, corresponding to the propagators of planar diagrams, as
\begin{equation}\label{kkplanars}
    \tilde{X}_{i,j} \coloneq X_{i,j} - m^2_{i,j} \ .
\end{equation}
As before, the values of these variables vanish on the boundary of the mesh: $\tilde{X}_{i,i+1} = 0, \ 
\tilde{X}_{1,n} = 0$. The tilded non-planar variables from Eq.~\eqref{tilde X c equation} are no longer equal to the massless ones, but instead
\begin{equation}\label{kktildecnonplanar}
\tilde{c}_{i,j} = c_{i,j} + 2m^2 n_i n_j \ .
\end{equation}
Using the on-shell condition $p_i^2 = m_i^2 = m^2 n_i^2$, we observe that we can write Eq.~(\ref{kktildecnonplanar}) as 
\begin{align}
    \tilde{c}_{i,j} &= - 2 p_i \cdot p_j + m^2 n_i n_j \nonumber \\
    &= - [ (p_i + p_j)^2 - m^2 (n_i + n_j)^2 ] = - \tilde{s}_{ij} \ ,
\end{align}
where $\tilde{s}_{ij} = s_{ij} - m_{i,j}^2$, making evident that $\tilde{c}_{i,j}$ is the usual massless non-planar variable but in $d+1$-dimensions \cite{gonzález2022doublecopymassivescalar}. In the same way, the massive planar variables in Eq.~(\ref{kkplanars}) are also equivalent to massless planar variables in $d+1$ dimensions.

Because of the existence of a $c$-equation with the same structure as the massless theory, when written in terms of tilded variables it is immediate that the dimensional reduction of $\Tr \Phi^3$ has an analogous set of hidden zeros but now in terms of $\tilde{c}_{i,j}$. The factorisation pattern also takes a similar form:
\begin{equation}\label{KKfactorisation}
\begin{aligned}
  A^{\tx{KK}}_{n}
  &\xrightarrow{\tilde{c}_{*} \neq 0}
  \Bigl(
    \frac{1}{X_{T} - m_{T}^{2}}
    + \frac{1}{X_{B} - m_{B}^{2}}
  \Bigr)
  A^{\tx{KK,up}} A^{\tx{KK,down}}
  \\
  &= \frac{\tilde{c}_{*}}{\tilx_{T}\,\tilx_{B}}\,
     A^{\tx{KK,up}} A^{\tx{KK,down}} \,.
\end{aligned}
\end{equation}
Logically this follows from the fact that the theory is massless in $d+1$ dimensions; hidden zeros are known to hold in the massless theory in $d+1$ dimensions, which come as conditions on $\tilde{c}_{i,j}$ in amplitudes written in terms of $\tilde{X}_{i,j}$. The factorisation pattern of $d+1$-dimensional massless amplitudes around $\tilde{c}_{*} \neq 0$ is precisely as in Eq.~(\ref{generalfactorisation}). Taking a massless amplitude in $d+1$ dimensions that has been factorised around a zero, if we then decompose the $d+1$-dimensional massless planar variables into $d$- and $1$-dimensional components, as in Eq.~(\ref{kkplanars}), we arrive at the factorisation pattern in Eq.~(\ref{KKfactorisation}). 

%-----------------------------------
\section{Massive Non-linear Sigma Model}\label{Section: NLSM}

It has been shown that the zeros of the massless $\Tr \Phi^3$ theory are also zeros of the NLSM \cite{HZ}. Here, we investigate the faithfulness of these zeros in the NLSM under massive deformations. We consider the NLSM describing the dynamics of Goldstone bosons arising from the spontaneous breaking breaking of a global symmetry in the pattern $G_{L} \times G_{R} \rightarrow G$, where $G_{L}$ and $G_{R}$ are compact Lie groups. We focus on the case in which $G_{L} = G_{R} = G = U(N)$. The Lagrangian for the NLSM can be found using the coset construction \cite{Callan:1969sn}, which provides the building blocks out of which the effective field theory is constructed. At leading-order, the Lagrangian for the Goldstone modes is
\begin{equation}\label{NLSM Lagrangian}
    \lag_{\tx{NLSM}} = \frac{F^{2}}{4} \Tr \left(  \partial_{\mu} U \partial^{\mu} U^{\dagger} \right),
\end{equation}
where $U$ is an element of $U(N)$ and $F$ is the symmetry breaking scale. One can parameterise the $U(N)$ element in many ways, equivalent to choosing coordinates on the coset target manifold. At tree-level, all the zeros of $\Tr \Phi^3$ are shared by the NLSM. 

\subsection{Simple Massive NLSM}\label{NLSMwithmass}

As a starting point, we could add to Eq.~(\ref{NLSM Lagrangian}) a simple, uniform mass term for the pions as
\begin{equation}
    \lag_{\tx{mass}} = -  m^2 \Tr (\pi^2)/ 2 \ .
\end{equation}
Implicit in this expression is a choice of coset coordinates $\pi^{a}$; the mass term is not coset-coordinate independent, and so is only meaningful when defined in reference to a parameterisation of $U(x)$. Changing parameterisation will genereically transform the mass term into an infinite tower of polynomial interactions. Since this mass term has the sole effect of changing the on-shell condition to $p_{i}^2 = m^2$, but not the derivative interactions, it generically spoils any hidden zeros observed in the massless theory. For example, if we define $\lag_{\tx{NLSM}}+\lag_{\tx{mass}}$ using the exponential parametrisation, 
\begin{equation} 
    U = \exp \left( \sqrt{2} \frac{i}{F} \pi \right),
  \end{equation}
at four points we find
\begin{equation}
    A[1234] = \frac{1}{2F^2} \left( c_{1,3} - \frac{2}{3}m^2 \right). 
\end{equation}
The derivative interactions between the Goldstones now source the extra mass terms in the amplitudes. If we want to change just these constant contributions, then we require non-derivative interactions. In principle, one could bootstrap the potential terms necessary to maintain the hidden zeros, but instead, we will use arguments from symmetry to describe them.

\subsection{Massive NLSM through Spurions}\label{NLSMwithmassSPurion}

We want to give the Goldstones mass in a symmetry-controlled way. The standard way to do this is through a spurion added to the NLSM Lagrangian. Assuming a uniform mass spectrum, we construct the lowest order term involving a single explicit symmetry-breaking parameter that is invariant under the full global $U(N)_{L} \times U(N)_{R}$ of the original model when we allow the symmetry-breaking parameter to transform under the group action, that is, the parameter is a spurion, essentially introducing an extra building block to the coset construction. For a particular transformation rule this constrains the ways in which the parameter enters into the Lagrangian. After the desired terms have been written down, we set the spurion equal to its constant value. 

We take the spurion, $\mathbf{M}$, to transform as\footnote{The transformation property of the spurion  can be motivated from the theory's UV completion. For chiral perturbation theory, the UV completion is QCD with massive quarks. Quark masses break the chiral symmetry but it can be restored by letting the quark mass matrix transform as a spurion.} $\mathbf{M} \rightarrow L \mathbf{M} R^{\dagger}$, where $L \in U(N)_{L}$ and $R \in U(N)_{R}$. The leading-order term we can construct is thus \cite{Schwartz:2014sze, TongGaugeTheory, Gasser:1983yg, Penco:2020kvy}
\begin{equation}\label{Lagrangianpotfinal}
    \lag_{\tx{pot}} = \kappa \Tr (U \mathbf{M} + U^{\dagger} \mathbf{M}^{\dagger}) \ ,
\end{equation}
with $\kappa$ a constant to be determined. We can now set the spurion to be a true constant, choosing a uniform mass spectrum so that $\mathbf{M} = m^2 \mathds{1}$. We choose to work within the exponential parameterisation, set  $\kappa =F^2/4$ to have a canonical mass term, and subtract $2 \kappa m^2 N$ to remove the constant vacuum term arising from Eq.~\eqref{Lagrangianpotfinal}. Thus, the potential term we consider is \begin{equation}\label{massivenlsmpotfinalfinal}
    \lag_{\tx{pot}} = \frac{F^2 m^2 }{4}\Tr (U + U^{\dagger} - 2) \ .
\end{equation} 
We will refer to the theory with the Lagrangian given by $\lag_{\tx{NLSM}}+\lag_{\tx{pot}}$ as mNLSM.

\subsubsection{Hidden Zeros}

We now proceed to look at the amplitudes of this theory, starting at four points. In the exponential parameterisation the four-point vertex from the potential term is
\begin{equation}
    V_{4}^{\tx{mass}}[1234] = \frac{m^2}{3 F^2} \ ,
\end{equation}
and hence, the four-point amplitude for mNLSM reads
\begin{equation}\label{fourpointnlsmcorrectzeros}
    A_{4}^{\tx{mNLSM}}[1234] = - \frac{1}{2 F^2} (s_{13} - 2m^2)= \frac{c_{1,3}}{2F^2} \ , 
\end{equation}
where $c_{1,3} = \tilde{X}_{1,3} + \tilde{X}_{2,4}$. The tilde planar variables are defined as in Eq.~\eqref{Massive Planar Variable Definition} throughout this section. This is just the massless amplitude but with tilded planar variables, thus it has the same zero as the four-point massive $\Tr \Phi^3$ amplitude. 

As we will show below, the potential we have added to generate a mass term has the effect of simply replacing $X_{i,j}$ by $\tilde{X}_{i,j}$ in the NLSM amplitude. Then, the same argument as before applies: the tilded variables satisfy the same $c$-equations as the massless planar variables, and so the hidden zeros are always inherited. To prove this, we will use the minimal parametrisation \cite{Ellis:1970nt,Kampf_2013} 
\begin{align}\label{minparamU}
        U &= \sqrt{2} \frac{i}{F} \pi + \sqrt{1 - \frac{2 \pi^2}{F^2}} \nonumber \\
        &= 1 + \sqrt{2}\,\frac{i}{F}\,\pi \;-\; 2 \sum_{k=1}^{\infty} \left( \frac{1}{2F^{2}} \right)^{k} C_{k-1}\,\pi^{2k} \ ,
 \end{align}
where $C_{k}=\tfrac{1}{k+1}\binom{2k}{k}$ are the Catalan numbers. The colour-ordered vertex in this parametrization is 
\begin{equation}
    V_{2n+2}
= \left(\frac{1}{2F^{2}}\right)^{\! n}\,\frac{1}{2}
\sum_{k=0}^{n-1} C_{k}\,C_{\,n-k-1}
\sum_{i=1}^{2n+2} X_{i,\,i+2k + 2} \ .
\end{equation}
Writing this in terms of $\tilde{X}_{i,j}$ we have 
\begin{equation}
        V_{2n+2}
= \left(\frac{1}{2F^{2}}\right)^{\! n}\,\frac{1}{2}
\sum_{k=0}^{n-1} C_{k}\,C_{n-k-1}
\sum_{i=1}^{2n+2} \left( \tilde{X}_{i,\,i+2k + 2} + m^2 \right),
\end{equation}
and we can isolate and simplify the constant term to find 
\begin{equation}
    \frac{1}{2^{n+1} F^{2n}} (2n+2)m^2 \, \sum_{k=0}^{n-1} C_{k} C_{n-k-1}=  \frac{(2n+2)m^2}{2^{n+1} F^{2n}} \, C_{n} \ ,
\end{equation}
where in the last equality we used the recurrence relation for the Catalan numbers. Now, looking at the potential term in the minimal parameterisation we have 
\begin{equation}
    \frac{m^2 F^2}{4}\Tr( U + U^{\dagger} - 2) = - \sum_{r=1}^{\infty} \frac{m^2}{2^{r} F^{2r - 2}} C_{r-1} \Tr(\pi^{2r}) \ .
\end{equation}
Isolating the coefficient of $\pi^{2n + 2}$ with $r = n+1$ gives  $- m^2C_{n}/(2^{n+1} F^{2n})$. To get the combinatorics right for the colour-ordered vertex we must multiply this by the number of elements in each cyclic equivalence class at that order, which is just $(2n + 2)$, giving us the general mass vertex contribution 
\begin{equation}
   V_{2n + 2}^{\tx{mass}} = - \frac{(2n+2)m^2}{2^{n+1} F^{2n}} C_{n} \ .
\end{equation}
In the sum we have 
\begin{align}
    &\tilde{V}_{2n+2} = V_{2n+2} + V_{2n+2}^{\tx{mass}} = \nonumber \\ &
    \left(\frac{1}{2F^{2}}\right)^{\! n}\,\frac{1}{2}
\sum_{k=0}^{n-1} C_{k}\,C_{\,n-k-1}
\sum_{i=1}^{2n+2} \left( \tilde{X}_{i,\,i+2k + 2} \right),
\end{align}
which is just the replacement $X \rightarrow \tilde{X}$ in the original NLSM vertex.  Thus, when constructing amplitudes for mNLSM, all the Feynman rules are those of the massless case with the replacement $X \rightarrow \tilde{X}$. Since the choice of parametrisation does not affect any statement about amplitudes and the $c$-variables remain unchanged (see Eq.~\eqref{tilde X c equation}), the hidden zeros of massive $\Tr \Phi^3$ are inherited by mNLSM to all orders at tree-level. 

%-----------------------------------
\subsubsection{Factorisation}
When the up and down amplitudes are even-point, NLSM amplitudes factorise in precisely the same pattern as $\Tr \Phi^3$, as in Eq.~(\ref{generalfactorisation}). If they are odd-point then the factorisation pattern is as in Eq.~(\ref{odd-point NLSM factorisation}), comprised of amplitudes coming from an extended theory of pions coupled to bi-adjoint cubic scalars, dubbed $\text{NLSM} + \phi^3$. This theory is found in the sub-leading soft limit of the NLSM \cite{Cachazo_2016_cubicandnlsm}, or as the low-energy limit of semi-abelian Z-theory \cite{Carrasco:2016ygv}. The bi-adjoint scalars are charged under $SU(N) \otimes SU(\tilde{N})$ with generators $T^{a} \otimes \tilde{T}^{\tilde{a}}$; the pions transform trivially under $SU(\tilde{N})$, with generators $T^{a} \otimes \mathds{1}$. Thus, for any process in this theory involving $r$ bi-adjoint scalars and $n-r$ pions, the amplitudes decompose with double colour ordering as
\begin{multline}
    \mathcal{A}_{n,r}^{\text{NLSM} + \phi^3} = \sum_{\sigma \in S_{r-1}} \sum_{\tau \in S_{n-1}} \, \Tr(T^{1} T^{\tau(2)} \dots T^{\tau(n)})  \, \times \\ \Tr(\tilde{T}^{1} \tilde{T}^{\sigma(2)} \dots \tilde{T}^{\sigma(r)}) \, A^{\text{NLSM} + \phi^3} [1, \tau | 1, \sigma],
\end{multline}
where $\tau$ and $\sigma$ are the orderings on the $SU(N)$ and $SU(\tilde{N})$ indices, respectively.

This theory has four types of vertices entering at leading order at tree level: the usual NLSM vertices between pions, a cubic vertex between the bi-adjoint scalars, even-point derivative vertices containing two $\phi$ and $2n-2$ pions, and non-derivative odd-point vertices containing three $\phi$ and $n-3$ pions. The even-point mixed vertices are exactly the same as the NLSM vertices of the same multiplicity, and the odd-point mixed vertices in exponential parametrisation are given by  
\begin{multline}
    V_{2k+1}[1,2,\dots,2k+1 \, | \, 1, 2k+1, j] = \\ 
    \frac{i}{2 \sqrt{2}} \frac{- (-2)^{k} \lambda}{(2k+1)! F^{2k-2}} \left[ \binom{2k}{j-1} (-1)^{j-1} - 1 \right].
\end{multline}
The odd-point mixed vertices have at least two $\phi$ legs adjacent in the colour ordering. This theory is described by the Lagrangian \cite{Yin:2018hht} 
 \begin{multline}\label{NLSMphi3 Lagrangian}
    \lag_{\tx{NLSM}+\phi^3} = \frac{F^{2}}{4(N^2-1)} \Tr \left(  \partial_{\mu} U(\Psi^{+}) \partial^{\mu} U^{\dagger}(\Psi^{+}) \right) + \\ C \, \Tr \, \left(
        \bigl[ U(\Psi^{+}) - U(\pi) \bigr]
        \bigl[ U(\Psi^{-}) - U(-\pi) \bigr] \right) + \text{h.c.} ) + \\ \frac{\lambda}{3!} f^{abc} f^{\tilde{a} \tilde{b} \tilde{c}} \phi^{a \tilde{a}} \phi^{b \tilde{b}} \phi^{c \tilde{c}} \ ,
\end{multline}
where $C = \frac{6i F^3 \lambda}{(N^2 - 1)^{3/2}}$ and 
\begin{equation}
\Psi^{\pm}= \pi^{c} (T^{c})_{ab} \delta_{\tilde{a}\tilde{b}}\pm \sqrt{N^2 - 1} \, \phi^{c\tilde{c}} (T^{c})_{ab} (T^{\tilde{c}})_{\tilde{a}\tilde{b}},
\end{equation}
with $U(x)$ satisfying $U(x)^{-1} = U(-x)$ and $U(0) = 1$. The first term in Eq.~(\ref{NLSMphi3 Lagrangian}) sources the NLSM-type vertices and the second term the non-derivative odd-point vertices. This Lagrangian generates more vertices than the ones that enter into the $\text{NLSM} + \phi^3$ theory, but any $\text{NLSM} + \phi^3$ amplitude that arises from a hidden zero factorisation of the NLSM has strictly only three external bi-adjoint scalars and $2n - 4$ external pions, schematically in the form $A[\alpha \, | \, i \, j \, n]$. Thus, at tree level, no vertex with $n>3$ bi-adjoint scalars can appear and any amplitude we are interested in is described by the Lagrangian in Eq.~(\ref{NLSMphi3 Lagrangian})\footnote{In \cite{Yin:2018hht} the author showed that for a general configuration of external states the vertices with $n>3$ bi-adjoint scalars only arise as higher-order corrections to tree-level amplitudes, so these vertices would still play no role if we had more than three external bi-adjoint scalars.}. 

We proceed to show how factorisation near a hidden zero works for the mNLSM. It is not hard to see that again the up and down amplitudes will be as in the massless case but with $X \rightarrow \tilde{X}$, in both the even- and odd-point cases. In the even-point case, these are just again mNLSM amplitudes. For example, for the blue square zero corresponding to the six-point NLSM as shown in Fig.~\ref{fig:mesh-main}, if we take $c_{*} = c_{3,5}$ then the mNLSM amplitude factorises as 
\begin{equation}
    A_6 \xrightarrow{c_{3,5} \neq 0} \left( \frac{c_{3,5}}{\tilx_{1,4} \tilx_{2,5}} \right) (\tilx_{1,3} + \tilx_{2,4})(\tilx_{1,5} + \tilx_{2,6}).
\end{equation}
The same logic applies to odd-point factorisation, with the amplitudes taking the same form as the massless case with $X \rightarrow \tilde{X}$, but now these amplitudes come from some massive deformation of $\text{NLSM} + \phi^3$. For example, taking the six-point mNLSM amplitude on the green skinny rectangle zero in Fig.~\ref{fig:mesh-main} with $c_{*} = c_{1,3}$ the amplitude factorises as 
\begin{equation}\label{massive odd point factorisation}
    A_{6} \xrightarrow{c_{1,3} \neq 0} c_{1,3} \left( \frac{\tilx_{3,5} + \tilx_{4,6}}{\tilx_{3,6}} + \frac{\tilx_{1,5} + \tilx_{4,6}}{\tilx_{1,4}} - 1 \right).
\end{equation}
So not only are the pions and bi-adjoint scalars are massive, but the derivative NLSM-type vertices receive analogous corrections to what we saw for the mNLSM. Diagrammatically, the massive five-point $\text{NLSM} + \phi^3$ amplitude in Eq.~(\ref{massive odd point factorisation}) corresponds to 
\begin{equation}
\;
\tikz[baseline=-0.6ex,scale=0.55,every node/.style={font=\scriptsize}]{
  \coordinate (v) at (0,0);
  \draw[dashed]  (v) -- (-1.10, 0.00) node[left]  {3};
  \draw[solid]  (v) -- (-0.80, 0.80) node[left]  {4};
  \draw[dashed] (v) -- (-0.80,-0.80) node[left]  {2};
  \draw[solid] (v) -- ( 0.80, 0.80) node[right] {5};
  \draw[dashed] (v) -- ( 0.80,-0.80) node[right] {6};
}
\;+\;
\tikz[baseline=-0.6ex,scale=0.55,every node/.style={font=\scriptsize}]{
  \coordinate (L) at (0,0);
  \coordinate (R) at (-1.35,0);

  \draw[dashed] (L) -- (R);

  \draw[solid] (L) -- ( 1.10, 0.00) node[right] {5};
  \draw[solid]  (L) -- ( 0.80, 0.80) node[right] {4};
  \draw[dashed]  (L) -- ( 0.80,-0.80) node[right] {6};

  \draw[dashed]  (R) -- (-2.15, 0.80) node[left]  {3};
  \draw[dashed] (R) -- (-2.15,-0.80) node[left]  {2};
}
\;+\;
\tikz[baseline=-0.6ex,scale=0.55,every node/.style={font=\scriptsize}]{
  \coordinate (L) at (0,0);
  \coordinate (R) at (-1.35,0);

  \draw[dashed] (L) -- (R);

  \draw[solid] (L) -- ( 1.10, 0.00) node[right] {4};
  \draw[dashed]  (L) -- ( 0.80, 0.80) node[right] {3};
  \draw[solid]  (L) -- ( 0.80,-0.80) node[right] {5};

  \draw[dashed]  (R) -- (-2.15, 0.80) node[left]  {2};
  \draw[dashed] (R) -- (-2.15,-0.80) node[left]  {6};
}
\;,
\end{equation}
where the dashed lines are bi-adjoint scalars and the solid lines pions, with the kinematic replacements $\tilde{X}_{2,4} \rightarrow \tilde{X}_{1,4}$ and $\tilde{X}_{2,5} \rightarrow \tilde{X}_{1,5}$. While the non-derivative odd-point vertices do not receive corrections, the even-point vertices containing two $\phi$ and $2n-2$ pions do. We find that the potential that deforms the pure NLSM pion vertices and the even-point mixed vertices in the required way, without introducing other undesired vertex corrections, is
\begin{multline}
\label{eq:pot_NLSM_BAS}
    \mathcal{L}_{\tx{pot}} = \frac{F^2 m^2}{8 (N^2 - 1)} \Tr \, \bigl( U(\Psi^{+}) + U^{\dagger}(\Psi^{+}) + \\ U(\Psi^{-}) + U^{\dagger}(\Psi^{-}) - 4 \bigr) \, ,
\end{multline}
where we have coupled to the same spurion $\mathbf{M}$ as before and set it to its constant value. This potential gives a mass $m$ to the pions and bi-adjoint scalars, and has a $\mathbb{Z}_2$ symmetry. The NLSM and mixed NLSM-type vertices with two $\phi$'s both receive the same corrections from this potential, which can be obtained from the massless case by simply putting a tilde on all planar variables. Thus, the odd-point amplitudes arising in the factorisation of the mNLSM come from the Lagrangian $\lag_{\tx{NLSM}+\phi^3}+\mathcal{L}_{\tx{pot}}$ with $\mathcal{L}_{\tx{pot}}$ given as in Eq.~\eqref{eq:pot_NLSM_BAS}.

\subsubsection{Double Copy}
Hidden zeros can be shown to be a result of BCJ relations in double-copyable theories such as NLSM and YM \cite{zerosfromdoublecopy}. Given the existence of massive hidden zeros, one could also ask whether these theories satisfy BCJ relations, thus explaining the appearance of the zeros and allowing for a physical double copy. The double copy for the mNLSM is expected to give a physical theory only at four points and to develop unphysical poles at higher multiplicities since it does not satisfy spectral conditions \cite{Momeni:2020vvr, Johnson:2020pny}. These unphysical poles arise from inverting full-rank matrices, contrary to the lower-rank ones arising in the massless case. Nevertheless, it is possible that the mNLSM amplitudes conspire to cancel these poles, as in the topologically massive case \cite{Gonzalez:2021bes}. An intriguing observation is that the mNLSM amplitudes satisfy 4-point, BCJ-like relations obtained by taking the standard massless BCJ ones and performing the replacements $s_{ij}\rightarrow  c_{i,j}=s_{ij}-2m^2$, which are implied by the tilde replacements on planar variables for different colour orderings:
\begin{equation} 
\label{eq:bcj}
\frac{s_{13} - 2m^2}{s_{12} - 2m^2} A[1324] =A[1234] \ .
\end{equation} 
We highlight that these are not related to color-kinematics duality in the standard way. One should also note that mNLSM amplitudes do not satisfy the Kleiss-Kuijf relations, since the potential term contains fully symmetric colour structures. Nevertheless, Eq.~\eqref{eq:bcj} applied to different colour orderings can reduce the partial amplitudes basis to a single amplitude. Although it is not entirely justified, we naively follow \cite{Johnson:2020pny} and construct the four-point KLT double copy for the mNLSM, which gives
\begin{equation} 
\label{eq:mNLSM_DC}
\mathcal{M}_4^\text{DC}=\frac{1}{\Lambda^6}(s_{12}-2m^2)(s_{13}-2m^2)(s_{23}-2m^2) \ .
\end{equation}
This corresponds to the massless double copy with the replacements $s_{ij}\rightarrow s_{ij}-2m^2$. The amplitude in Eq.~\eqref{eq:mNLSM_DC} arises from the massive Special Galileon \cite{Cachazo:2014xea, Hinterbichler:2015pqa} plus contributions from additional derivative and non-derivative contact term interactions. We note that double copy constructions for theories with general colour building blocks have been explored in \cite{Carrasco:2019yyn,Low:2019wuv,Low:2020ubn,Carrasco:2021ptp,Carrasco:2022jxn,Bonnefoy:2021qgu,Carrasco:2026hxf} via compositions of these blocks, but we defer a complete analysis of the mNLSM double copy to future work.

\subsubsection{On-Shell Recursion}

On-shell recursion for NLSM amplitudes typically leverage the theory's enhanced soft behaviour through the Adler zero \cite{Cheung:2015ota}. For mNLSM we no longer have the Adler zero, instead having in the leading-order soft limit some non-vanishing constant controlled by the symmetry-breaking parameter $m$. The precise form of this ``soft" limit is dictated by the partially-conserved axial current (PCAC) \cite{Adler, Nambu:1960xd, Gell-Mann:1960mvl,Gasser:1983yg,Leutwyler:1993iq,Scherer:2005ri,Weinberg_PhysRevLett.17.616}. As a consequence, soft-recursion is not applicable to mNLSM. 

A new recursion relation for NLSM amplitudes based on hidden zeros has been proposed \cite{Li:2025suo}. Given that mNLSM amplitudes have the same hidden zeros as the massless theory, we may expect that this recursive procedure be applicable where soft recursion is not. In what follows we will reformulate this recursion relation in the language of planar/non-planar variables and apply it to mNLSM amplitudes. The procedure is applicable to NLSM amplitudes also, with massless planar variables substituted where appropriate. 

We first shift the planar and non-planar variables separately by a complex parameter $z$ as 
\begin{subequations}
\label{nonplanarshift}
\begin{align}
 c_{i,j}(z) &= c_{i,j} + z r_{i,j}, \\
    \tilde{X}_{i,j}(z) &= \tilde{X}_{i,j} + z g_{i,j},
\end{align}
\end{subequations}
where for a $2n$-point process we have introduced $n(2n-3)$ non-planar shift functions $r_{i,j}$ and $n(2n-3)$ planar shift functions $g_{i,j}$. To ensure that the shifted kinematic data are physical, we require that they satisfy the same momentum conservation relations as the unshifted variables, that is, the $c$-equations. This implies the $n(2n-3)$ relations
\begin{equation}\label{shiftcequation}
    r_{i,j} = g_{i,j} + g_{i+1, j+1} - g_{i+1,j} - g_{i, j+1} \ ,
\end{equation}
which are expected since we started with an over-complete kinematic basis. Under these complex shifts, mNLSM amplitudes are linear in $z$ at large $z$, so the familiar BCFW integrand $A(z) / z$ has a non-zero boundary term coming from the residue at $z = \infty$. If we instead consider the contour integral \cite{Cachazo:2021wsz, Li:2025suo}
\begin{equation}\label{HZrecursionfunction}
    \frac{1}{2 \pi i} \oint \, dz \, \frac{A^{\tx{mNLSM}}_{2n}(z)}{z (1-z^2)}
\end{equation}
then we have good UV behaviour and can express the mNLSM amplitude as 
\begin{equation}\label{eq:HZrecursion}
    A^{\tx{mNLSM}}_{2n} = \!\!\!\!\!\!\sum_{z^{*} : \tilde{X}(z^{*}) = 0} \!\!\!\!\!\!A^{\text{mNLSM}}_{L}(z^*) \frac{1}{\bigl[1 - (z^*)^2\bigr] \tilde{X}} A^{\text{mNLSM}}_{R}(z^*) ,
\end{equation}
where $\tilde{X}$ is an unshifted massive planar variable corresponding to a physical factorisation channel. Following \cite{Li:2025suo}, we have removed two unwanted terms coming from the poles at $z = \pm 1$ with a judicious choice of shifts in Eq.~(\ref{nonplanarshift}) that guarantee that the residue of these two poles is proportional to the $2n$-point mNLSM amplitude evaluated on a hidden zero. This ensures a good recursion with no boundary term, formulated only in terms of lower-point amplitudes. 

There is a subtlety in how we choose which hidden zeros to use. Not only do we require two distinct hidden zeros for the two unwanted pole contributions, but these hidden zeros may not set to zero any non-planar variable shared by the other. In other words, on the kinematic mesh they must be two zeros that come from non-overlapping causal diamonds. If we consider the full set of planar and non-planar shifts $S = \{r_{i,j} \, , \, g_{i,j} \}$, we see it has size $|S| = 2n(2n-3)$. For two skinny rectangle zeros we set to zero $2(2n - 3)$ non-planar variables, determining $2(2n-3)$ of the $r_{i,j}$ in $S$. We also have $n(2n-3)$ $c$-equations, after which we are left with $(n-2)(2n-3)$ undetermined variables in $S$. For $2n=6$, this amounts to three undetermined degrees of freedom in $S$. The final recursion relation in Eq.~(\ref{eq:HZrecursion}) must be independent of the undetermined variables, so we would like to set them equal to zero for simplicity. Naively, we could choose any configuration of $(n-2)(2n-3)$ variables in $S$ to set to zero in order to fully determine the system in terms of fixed values, but there are further subtleties in how we can choose these variables so as not to create inconsistencies. We explain this in detail in Appendix \ref{app:variable_shifts}. An important condition is that if we are choosing to set multiple $r_{i,j}$ to zero, then there is a consistency requirement (Eq.~\eqref{eq:consistency} we must ensure is upheld.

It is worth noting that one can formulate an equivalent recursion for mNLSM in terms of Mandelstams, as was originally done for the NLSM in \cite{Li:2025suo}. But here, a hidden zero shift would take the form, $r_{1,3} = s_{13} - 2m^2$, and the simplicity of constructing the recursion in terms of planar data is lost. Formulating the above in terms of planar data means that the unchanged form of the $c$-equations results in a recursive procedure that works the same for NLSM and mNLSM, simply with $X \leftrightarrow \tilde{X}$, and we do not have to contend with an increasing number of $m^2$ terms. In Appendix~\ref{app:mNLSMrecursion} we calculate $A^{\tx{mNLSM}}_{6}$ explicitly using the recursion relation in Eq.~\ref{eq:HZrecursion}. 

\subsection{Dimensional Reduction of NLSM}
Similarly to $\Tr \Phi^3$, we can also look at the KK reduction of the NLSM. If we take the spacetime to be the same $\mathds{R}^{d-1,1} \times S^{1}$ as before and expand the pions as $\pi(x^{\mu},y) = \sum_{n=-\infty}^{\infty} \pi_{n} (x^{\mu}) e^{inmy}$,
where $\pi = \pi^{a} T^{a}$ and $\pi_n = \pi^{a}_n T^{a}$, then the quadratic part of the Lagrangian in Eq.~(\ref{NLSM Lagrangian}) is 
\begin{equation}
    \lag^{(2)} = \frac{1}{2} \sum_{n} \left( \partial_{\mu} \pi_{n} \partial^{\mu} \pi_{n} - n^2 m^2 \pi_n^2 \right).
\end{equation}
The pions now have a mass that depends on their mode number, as we had with KK $\Tr \Phi^3$. The dimensional reduction has the effect of changing $X \rightarrow \tilde{X}$  on the interactions \cite{gonzález2022doublecopymassivescalar}, where now the planar variables are those defined in Eq.~\eqref{kkplanars}. The mode number is again conserved at each vertex and $\sum_{i} n_i = 0$ for each diagram.

As an example, the four-point amplitude in this theory is simply 
\begin{equation}\label{KKNLSMfourpointexample}
    \frac{1}{2F^2} (\tilde{X}_{1,3} + \tilde{X}_{2,4}) = - \frac{1}{2F^2} \tilde{c}_{1,3}=\frac{1}{2F^2}(  s_{ij} - m_{i,j}^2) \ .
\end{equation}
We see that the KK NLSM shares the same four-point $\tilde{c}_{1,3} = 0$ hidden zero as KK $\Tr \Phi^3$. This persists at all multiplicity, and all zeros of KK $\Tr \Phi^3$ are zeros of KK NLSM.

As a result we can expect the amplitudes of this theory to factorise around hidden zeros in manner consistent with what we have already seen; the amplitudes upon factorisation will look like those of the massless theory but with the replacements $X \rightarrow \tilde{X}$. The mixed theory providing the odd-point sub-amplitudes in this case is the dimensional reduction of the Lagrangian in Eq.~(\ref{NLSMphi3 Lagrangian}) with pions expanded in terms of Fourier modes as before. 

%-----------------------------------
\section{Massive Yang-Mills}

YM amplitudes have zeros at the same loci in kinematic space as $\Tr \Phi^3$ and the NLSM, if additional conditions on the polarisation vectors are met. If these conditions are not met in addition to the scalar zero conditions then the zero is not present, and instead they factorise into a sum of lower-point amplitudes \cite{Guevara:2024nxd,Zhang:2024efe}. 

Below we analyse whether an amplitude involving massive gluons can share the hidden zeros of its massless counterpart, like we found in the NLSM. As in the NLSM, we expect massive hidden zeros to be present in a theory where the mass is added in a way that takes the symmetries of the model into account. For massive gauge bosons the immediate candidate is a YM-Higgs theory with spontaneously broken gauge symmetry, but as in the NLSM case, we analyse both the naive and controlled massive deformations.

%-----------------------------------
\subsection{Massive YM Theory}
We consider a massive deformation of the YM Lagrangian given by
\begin{equation}
    \lag = - \frac{1}{4} \Tr ( F_{\mu \nu} F^{\mu \nu}) - \frac{1}{2}m^2 \Tr (A_{\mu} A^{\mu}) \ , 
\end{equation}
which breaks the gauge symmetry of the theory\footnote{Note that gauge invariance can always be restored via the St\"{u}ckelberg trick \cite{Kunimasa:1967zza,Hinterbichler:2011tt}.}. It is straightforward to see that this theory does not exhibit hidden zeros in parallel with the other massive theories thus far considered. After enforcing the spin-one hidden zero conditions from Eq.~\eqref{spin1zeroconditions}, the four-point partial amplitude reduces to \cite{WardMScThesis2025HiddenZerosMassive}
\begin{align}
\label{leftoverbit}
    \mathcal{A}^{m\tx{YM}}_{4} [1234] \, \big|_{\tx{HZ}} = \frac{m^2}{\tilde{X}_{1,3}} [ & (\ep_1 \cdot \ep_2)(\ep_3 \cdot \ep_4) \nonumber \\
    & - (\ep_2 \cdot \ep_3)(\ep_1 \cdot \ep_4) ] \ .
\end{align}
At four points we have the identity $\tilx_{1,3} + \tilx_{2,4} = c_{1,3}$, so that on the four-point hidden zero locus we have $\tilx_{1,3} = - \tilx_{2,4}$. As a result, Eq.~(\ref{leftoverbit}) takes the form 
\begin{multline}
    \mathcal{A}^{m\tx{YM}}_{4} [1234] \, \big|_{\tx{HZ}} = m^2 \biggl[ \frac{(\ep_1 \cdot \ep_2)(\ep_3 \cdot \ep_4)}{\tilde{X}_{1,3}} + \\  \frac{(\ep_2 \cdot \ep_3)(\ep_1 \cdot \ep_4)}{\tilde{X}_{2,4}} \biggr] \ ,
\end{multline}
which looks conspicuously like the exchange of a scalar with mass $m$ between gluons at four points. We can see that the hidden zero can be recovered in the high energy limit $X_{i,j}\gg m^2$, as expected.

We should at this point highlight a source of possible contention. The spin-$1$ hidden zero conditions in Eq.~(\ref{spin1zeroconditions})--defined originally for massless vectors and now appropriated for massive ones--were defined as such due the gauge invariance of YM. Because of the gauge freedom in shifting $\epsilon^{\mu} \rightarrow \epsilon^{\mu} + \alpha p^{\mu}$ due to the Ward identity, conditions on the momenta and polarisations separately are not gauge invariant. To select a particular element of the gauge orbit we require $\epsilon_i \cdot p_j = 0$ and $\epsilon_j \cdot p_i = 0$ in addition to the other two conditions in Eq.~(\ref{spin1zeroconditions}). However, in the theory of massive vectors considered here, we no longer have this gauge invariance. This raises the question of whether the spin-$1$ hidden zero conditions in Eq.~(\ref{spin1zeroconditions}) should be applicable and what physical significance we ascribe to them. 

%-----------------------------------
\subsection{Spontaneously Broken YM Theory}
As in the case of the mNLSM, we need to consider a massive vector model where symmetries are broken in a controlled manner; hence, we look at a theory with spontaneously broken gauge symmetry. We consider a gauge field coupled to scalars, $\Phi=\Phi^aT^a$, in the adjoint:
\begin{equation}
    \mathcal{L}=-\frac{1}{2}\,\operatorname{tr} F_{\mu\nu}F^{\mu\nu}+\operatorname{tr} D_{\mu}\Phi D^{\mu}\Phi - V(\Phi) \ ,
\end{equation}
We remain agnostic of the specific symmetry breaking potential since we only care about the symmetry breaking pattern, which we choose to be $U(N)\rightarrow U(1)^{N}$, and which can be obtained by considering a diagonal vacuum expectation value (vev) for the scalars $\braket{\Phi}=\text{diag.}(v_1,v_2,\cdots,v_N)$. This gives a mass $m_{i,j}=g|v_i-v_j|$ to the off-diagonal gauge fields $A_i^j$, where $i,j$ are indices in the fundamental and anti-fundamental representation of $U(N)$ respectively. The interactions between gauge bosons couple $A_i^j$ and $A_j^k$ to $A_i^k$ at three-points and $A_i^j$, $A_j^k$, and $A_k^l$ to $A_i^l$ at four-points. The propagators will be given by the planar variables 
\begin{equation}\label{eq:SSBplanars}
    \tilde{X}_{i,j} \coloneq X_{i,j} - g^2( v_i-v_{j})^2 \ .
\end{equation}
Thus, the tilde non-planar variables can now be expressed as
\begin{align}
\tilde{c}_{i,j} &=c_{i,j} + 2g^2 (v_i-v_{i+1})(v_j-v_{j+1}) \nonumber \\
&=c_{i,j} + 2m_i^2m_j^2\ , \label{eq:SSBtildecnonplanar}
\end{align}
where $m_{i,j}$ is the mass if the $i/j$-th particle in the partial amplitude. We can see that this kinematic variables are equivalent to those arising in the KK reduction of the theories discussed above. In fact, it has been shown that the amplitudes can be recovered from the massless ones by taking  $X\rightarrow \tilde{X}$ and leaving all the dot products involving polarization vectors unchanged \cite{Naculich:2015coa}, which guarantees the existence of hidden zeros.

The amplitudes of this theory are intimately related to the dimensional reduction of the massless YM \cite{Naculich:2015coa}. If we consider $M\geq 1$ adjoint scalars acquiring a vev, the massive gluon amplitudes can be obtained from a dimensional reduction from $d+M$ to $d$. More precisely, the amplitudes are obtained by taking the $d+M$-dimensional momentum of a particle $i$ and splitting it as $K_{i} = (k_i , \kappa_{i})$,
where $k_{i}$ is the momentum in $d$-dimensions which now has mass $m_i = |\kappa_{i}|$ and choosing the $(d + M)$-dimensional  polarisation vectors for massless gauge bosons $\mathcal{E}_{i}$ to have the form $\mathcal{E}_{i} = (\epsilon_i , 0)$. This implies
\begin{subequations}
    \begin{align}
    K_{i} \cdot K_{j} &= k_{i} \cdot k_{j} - \kappa_{i} \cdot \kappa_{j}, \\
    K_{i} \cdot \mathcal{E}_{j} &= k_{i} \cdot \epsilon_{j}, \\ 
    \mathcal{E}_{i} \cdot \mathcal{E}_{j} &= \epsilon_{i} \cdot \epsilon_{j} \ .
\end{align}
\end{subequations}
Since only the momenta dot products are affected and there is still a notion of internal momentum conservation at vertices which gives rise to conditions on allowed exchange masses, the amplitudes can be obtained as those of the massless theory with the replacements
\begin{equation}\label{eq:DIMREDplanars}
X_{i,j}\rightarrow\tilde{X}_{i,j} \coloneq X_{i,j} -  (\kappa_i + \dots + \kappa_{j-1})^2 \ .
\end{equation}
from which we can recover the notion of hidden zeros.

%-----------------------------------
\section{Discussion}
We have shown that certain mass deformations allow the existence of hidden zeros and factorisations near them. We analysed simple mass deformations without additional potential terms or interactions and showed that the zeros only survive for the $\Tr \Phi^3$ theory. We also considered dimensionally reduced theories. Since the hidden zeros can be derived from the BCJ relations \cite{zerosfromdoublecopy}, which the KK theories satisfy, these properties should hold, as we showed above.

In the NLSM, a subset of the hidden zeros is known to imply the Adler zero \cite{Adler}. This occurs due to the absence of relevant poles, which is not the case in the $\Tr \Phi^3$ theory \cite{HZ}. For the massive deformations considered here, both the mNLSM and KK constructions introduce a breaking of the constant shift symmetry, so the Adler zero and the usual formulation of a recursion relation for the amplitudes are lost. We have shown here that is possible to extend the recent results of  \cite{Li:2025suo} and formulate a recursion relation for the mNLSM amplitudes by shifting appropriately a set of planar and non-planar variables. Other interesting properties of theories with hidden zeros, such as an improved UV behaviour \cite{Rodina_2025,Jones:2025rbv}, are also likely to break in the mNLSM. This leaves open the question of the physical significance of hidden zeros in massive theories. Another open question is whether a different choice of spurion mass matrix, instead of the one leading to the uniform spectrum, also preserves the hidden zeros.

The double copy of the mass-deformed theories which arise due to a spontaneous symmetry breaking / dimensional reduction that were considered above is known to give a physical theory \cite{gonzález2022doublecopymassivescalar,Johnson:2020pny,Momeni:2020hmc,Chiodaroli:2015rdg,Chiodaroli:2017ehv}. On the other hand, the mNLSM does not have a standard double copy construction due to the failure to satisfy KK relations, but the insights of \cite{Carrasco:2019yyn,Low:2019wuv,Low:2020ubn,Carrasco:2021ptp,Carrasco:2022jxn,Bonnefoy:2021qgu} can aid to formulate this.  A satisfactory generic massive double copy is, as yet, elusive. Generalising the massless procedure to massive theories is only generally valid at four points for non-dimensionally-reduced theories, and the formalism is qualitatively changed. In the massless case, hidden zeros are known to be directly derivable from BCJ relations, and propagate through the web of double-copy theories through the KLT formalism. Given that hidden zeros in massive theories are not connected always to the massless-to-massive double copy procedure, it is natural to ask whether they may inform us of some potentially new formulation, or whether they exist in isolation.

In the case of gauge theories, we observed that the failure of the hidden zero for a massive gluon at four-points has the form of a scalar interaction between gluons. This is the contribution from the longitudinal polarization that is present in the massive case and spoils the massless hidden zero. We also highlighted that once the gauge symmetry is broken, the standard YM hidden zero constraints are not justified. Last, we looked at theories with a spontaneously broken $SU(N)$, which straightforward satisfy hidden zeros. It would be interesting to prove if this holds more generally for other symmetry breaking patterns and different gauge groups. Additionally, we only considered the amplitudes with external gluons,  but one can look at the more general case where amplitudes involve both massive gluons and massive adjoint scalars as external states. Finally, another possible future direction is to understand how the factorizations of \cite{Zhang:2024efe,Guevara:2024nxd}, which simply impose the constraints from the scalar hidden zeros in Eq.~\eqref{eq:hz}, extend to theories involving massive gluons. For the cases arising from spontaneous symmetry breaking, some realisation of the factorization is expected to hold, but it would be desirable to see the explicit form in terms of amplitudes of gluons and adjoint scalars.\\

\noindent{{\bf{\em Acknowledgments.}}}
M.C.G is supported by the Imperial College Research Fellowship.

\appendix

\section{Choice of Shifts in Recursion Relation} \label{app:variable_shifts}

Here we analyse in detail the subtleties arising when choosing the shifts of the planar and non-planar variables for the recursion relation. 

Firstly, setting shifts to zero breaks the linear dependence of the variables in $z$. To ensure the UV behaviour is not spoiled, we must choose these shifts such that planar variables in propagator denominators of physical factorisation channels remain linear in z, as this ensures there is no worsening of the large-z behaviour. At $2n=6$ for the mNLSM this means we leave $g_{1,4}$, $g_{2,5}$, and $g_{3,6}$ alone. 

Secondly, the system of $c$-equations for even-point scattering has a one-dimensional kernel, meaning the $g_{i,j}$ are determined up to a one-parameter freedom\footnote{This is not the case for odd-point processes, where the $c$-equations have full rank. This freedom is equivalent to the $\delta$-shifts of \cite{HZ}.} \cite{HZ}. If we write the linear system of equations as 
\begin{equation}\label{matrix}
    r = \mathbf{A} g,
\end{equation}
where $r,g$ are $n(2n-3)$-dimensional vectors containing the non-planar and planar shifts respectively, then the matrix $\mathbf{A}$ has a single null vector, whose components are
\begin{equation}
    v_{i,j} =
  \begin{cases}
    (-1)^{i} & \text{if $j-i$ even}, \\
    0 & \text{if $j-i$ odd}.
  \end{cases} 
\end{equation}
For $2n=6$ with the index pair ordering $(13), (14), (15), (24), (25), (26), (35), (36), (46)$, this corresponds to the null vector 
\begin{equation}
    v = (-1, 0, -1, 1, 0, 1, -1, 0, 1)^{T}.
\end{equation}
This is both the left and right null vector of $\mathbf{A}$. This means that $\mathbf{A}$ is rank-deficient and the $g_{i,j}$ are defined only up to shifts proportional to a single parameter. Crucially, $g_{i,j}$ shifts with even-separated indices do not transform under this redundancy, and mNLSM propagators in physical factorisation channels are unaffected. Therefore, to fully specify a set of solutions, we must manually fix the value of a single $g_{i,j}$ within this redundancy class. We will always set this value to zero. Importantly, this does not use up one of our choices of undetermined variables in $S$.

\begin{figure}
    \centering
    \includegraphics[width=0.5\linewidth]{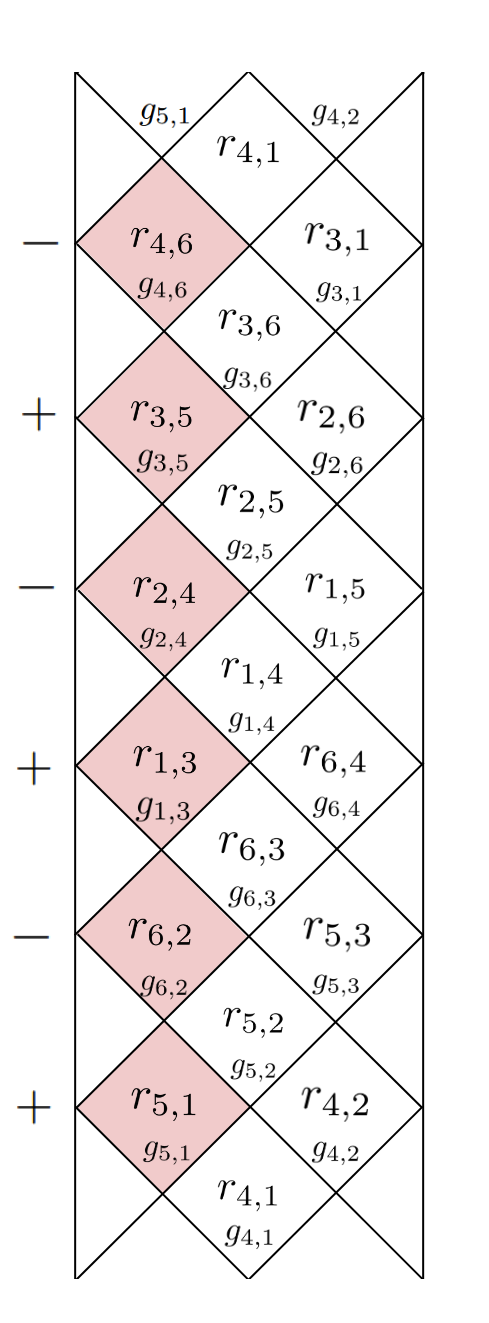}
    \caption{Summing the tiles in one cycle on any edge column of a kinematic mesh with alternating relative signs, shown here for $2n=6$, is a visual way to view the consistency requirement on $c_{i,j}$/$r_{i,j}$ that results from the one-dimensional kernel of the $c$-equations. The sum fully telescopes, leaving a relation solely in terms of $c_{i,j}$/$r_{i,j}$. }
    \label{fig:column}
\end{figure}

Thirdly, once we start fixing values of $r_{i,j}$ with hidden zeros, not all the non-planar shifts $r_{i,j}$ are actually independent. In fact, they are related to each other by virtue of the null space of the $c$-equations. If we act on Eq.~(\ref{matrix}) with the null vector $v$ we have 
\begin{equation}
    v^{T} r = 0. \label{eq:consistency}
\end{equation}
For $2n=6$ this gives
\begin{equation}
    r_{1,3} + r_{1,5} + r_{3,5} - r_{2,4} - r_{2,6} - r_{4,6} = 0 \ , 
\end{equation}
which, by virtue of the structure of the null vector of $\mathbf{A}$, is a relation between $r_{i,j}$ for which the difference between the values of their indices is even. This holds generally at any multiplicity. If we are solving the system of equations with $r_{i,j}$ as unknowns and $g_{i,j}$ as input data, this relation is just a guaranteed identity on the solutions $r_{i,j}$, but if we specify certain solutions $r_{a,b}$ beforehand, such as hidden zero shifts, then this relation becomes a consistency requirement on the remaining undetermined shifts that must be satisfied for the system to be soluble. In other words, it is the statement that $r \in \text{Im}(\mathbf{A})$. Because $\mathbf{A}$ has a one-dimensional kernel, $\text{Im}(\mathbf{A})$ has codimension one, and we thus have $(n-2)(2n-3) - 1$ undetermined variables. For example, with the hidden zeros $c_{1,3} = c_{1,4} = c_{1,5} = 0$ and $c_{2,4} = c_{2,5} = c_{2,6} = 0$, we have $r_{3,5}$, $r_{3,6}$, $r_{4,6}$ undetermined. Trying to naively set all three to zero results in $r_{1,3} + r_{1,5} - r_{2,4} - r_{2,6} = 0$, which is not true for our choice of shifts. We can visualise this relation on the mesh as being the unique fully telescoping column--valid for general $n$-point scattering--shown in Figure~(\ref{fig:column}). The substance of the above is that, if we are choosing to set multiple $r_{i,j}$ to zero, then there is a consistency requirement we must ensure is upheld. 

From the above, there are many permissible configurations of variables to set equal to zero. We could use all the freedom on the $r_{i,j}$s or all the freedom on the $g_{i,j}$s, or some admixture of the two. If choosing multiple $r_{i,j}$ then we need to bear the consistency requirement in mind. When setting $g_{i,j}$s to zero there are also situations we have to avoid, in addition to the factorisation channel requirement. As an example, take the same six-point hidden zero as before. If we set $g_{1,3} = g_{2,4} = 0$ then we generate the unphysical relation $r_{2,4} + r_{2,5} + r_{2,6} = c_{2,4} + c_{2,5} + c_{2,6} = 0$. There are multiple relations that can be generated like this depending on the choice and number of $g_{i,j}$ we would set to zero and the configuration of the hidden zeros. Any choice must obey these $r$- and $g$-consistency checks. 

Using these facts, we can look concretely at the $2n=6$ case, where we use the $c_{1,3} = c_{1,4} = c_{1,5} = 0$ and $c_{2,4} = c_{2,5} = c_{2,6} = 0$ hidden zeros. We naively have three undetermined variables to set equal to zero, but we know that one of the even-separated $r_{i,j}$ is dependent on the others. The odd-separated $r_{i,j}$ are all independent, so a valid choice of variables to set to zero generally would be: one valid $g_{i,j} = 0$ to fix the one-parameter null freedom, all remaining odd-separated $r_{i,j} = 0$, and all but one remaining even-separated $r_{i,j} = 0$, with the final one determined by the others. In the $2n=6$ case, this could correspond to $r_{3,6} = r_{4,6} = g_{2,6} = 0$. We could also choose all variables as $g_{i,j}$; such choices, if they do not induce spurious relations among the $c_{i,j}$'s, result in $r_{i,j}$ that automatically satisfy the $r$-consistency requirement. One example is $g_{4,6} = g_{1,5} = g_{2,5} = 0$. Note that trying to set all $g$-freedom to zero is difficult to do consistently. Without a rule for determining safe configurations, one could either set all the $r$-freedom to zero or set to zero some $r_{i,j}$ along with a selection of the $\sum^{2n-5}_{i} i$ number of $g_{i,j}$ that are not in the boundary of the hidden zero causal diamonds for two skinny rectangles. This number of $g_{i,j}$ not in contact with any hidden zero causal diamond is maximised by two skinny rectangle zeros.

\section{Hidden Zero On-Shell Recursion for mNLSM at Six Points} \label{app:mNLSMrecursion}

Here we construct an explicit example of the planar recursion relation described in Section~\ref{NLSMwithmassSPurion}. 

We choose the hidden zeros $c_{1,3} = c_{1,4} = c_{1,5} = 0$ and $c_{2,4} = c_{2,5} = c_{2,6} = 0$, fixing the following shifts: 
\begin{equation}
    \begin{aligned}
        r_{1,3} &= -c_{1,3}, \quad r_{1,4} = -c_{1,4}, \quad r_{1,5} = -c_{1,5}, \\
         r_{2,4} &= c_{2,4}, \quad r_{2,5} = c_{2,5}, \quad r_{2,6} = c_{2,6} \ .
    \end{aligned}
\end{equation}
We choose to eliminate the freedom in the shifts system by setting $r_{3,6} = r_{4,6} =g_{2,6}= 0$, from this, we determine $r_{3,5}$ using the $r$-consistency requirement as $r_{3,5} = c_{1,3} + c_{1,5} + c_{2,4} + c_{2,6}$. The full set of shifts can then be found to be given by
\begin{align}
r_{1,3}&=-c_{1,3}, \quad  r_{1,4}=-c_{1,4}, \quad  r_{1,5}=-c_{1,5}, \\
r_{2,4}&= -c_{2,4}, \quad  r_{2,5}=-c_{2,5}, \quad  r_{2,6}=-c_{2,6}, \\  
r_{3,6}&=0, \quad \quad \; \;  r_{4,6}=0, \\
r_{3,5}&=c_{1,3}+c_{1,5}+c_{2,4}+c_{2,6} \ ,
\end{align}
and 
\begin{align}
g_{1,3}&=-(c_{1,3}+c_{1,4}+c_{1,5}), \\
g_{1,4}&=c_{1,3}+c_{2,4}+c_{2,5}+c_{2,6}, \\
g_{1,5}&=c_{2,6}, \\
g_{2,4}&=c_{1,3}+c_{1,4}+c_{1,5}+c_{2,4}+c_{2,5}+c_{2,6}, \\
g_{2,5}&=c_{2,6}+c_{1,5}, \\
g_{2,6}&=0, \\
g_{3,5}&=-(c_{1,3}+c_{1,4}+c_{2,5}), \\
g_{3,6}&=-(c_{1,3}+c_{1,4}+c_{1,5}+c_{2,6}), \\
g_{4,6}&=c_{1,3}+c_{2,4}+c_{2,5} \  .
\end{align}
Focusing on the contribution from the pole at $\tilde{X}_{1,4}(z) = 0$  first, we have 
\begin{multline}
    \text{Res}_{\tilde{X}_{1,4}(z^{*})=0}  = \\\frac{[\tilde{X}_{1,3}(z^{*}) + \tilde{X}_{2,4}(z^{*})][\tilde{X}_{1,5}(z^{*}) + \tilde{X}_{4,6}(z^{*})]}{X_{1,4}[1 - (z^{*})^2]},
\end{multline}
with $z^{*} = - \tilde{X}_{1,4} / g_{1,4}$. We now have $c_{1,3}(z^{*}) = \tilde{X}_{1,3}(z^{*}) + \tilde{X}_{2,4}(z^{*}) - \tilde{X}_{1,4}(z^{*}) = \tilde{X}_{1,3}(z^{*}) + \tilde{X}_{2,4}(z^{*})$, so 
\begin{equation}
    \tilde{X}_{1,3}(z^{*}) + \tilde{X}_{2,4}(z^{*}) = (1-z^{*})c_{1,3} \ .
\end{equation}
Similarly, we have 
\begin{equation}
    \tilde{X}_{1,5}(z^{*}) + \tilde{X}_{4,6}(z^{*}) = c_{4,6} \  ,
\end{equation}
since $r_{4,6} = 0$,  so we can write this channel contribution as 
\begin{align}
    \text{Res}_{\tilde{X}_{1,4}(z^{*})=0}  &=\frac{c_{1,3} c_{4,6}}{\tilde{X}_{1,4}} \frac{1}{1+z^{*}} \\
    &= \frac{c_{1,3} c_{4,6}}{\tilde{X}_{1,4}} \frac{g_{1,4}}{g_{1,4} - \tilde{X}_{1,4}} \ .
\end{align}
Since $g_{1,4} = 2\tilde{X}_{1,3} + 2\tilde{X}_{2,4} - \tilde{X}_{1,4}$, we have that $g_{1,4} - \tilde{X}_{1,4} = 2c_{1,3}$, which cancels to give
\begin{equation}
    \text{Res}_{\tilde{X}_{1,4}(z^{*})=0}  = \frac{c_{4,6}}{\tilde{X}_{1,4}} \frac{g_{1,4}}{2} \ .
\end{equation}
Using $c_{4,6} = \tilde{X}_{4,6} + \tilde{X}_{1,5} - \tilde{X}_{1,4}$ gives the final contribution 
\begin{multline}
    \text{Res}_{\tilde{X}_{1,4}(z^{*})=0}  = \frac{(\tilde{X}_{1,3} + \tilde{X}_{2,4})(\tilde{X}_{1,5} + \tilde{X}_{4,6})}{\tilde{X}_{1,4}} \, + \\ \hf (\tilde{X}_{1,4} - \tilde{X}_{1,5} - \tilde{X}_{4,6} - 2 \tilde{X}_{1,3} - 2 \tilde{X}_{2,4}) \ .
\end{multline}
The contribution from the pole at $\tilde{X}_{2,5}(z) = 0$ is particularly simple and all $z$-dependence cancels out, leaving
\begin{equation}
    \text{Res}_{\tilde{X}_{2,5}(z^{*})=0}  = \frac{c_{1,5} \, c_{2,4}}{\tilde{X}_{2,5}}.
\end{equation}
Substituting in the $c$-equations for $c_{2,4}$ and $c_{1,5}$ gives the contribution 

\begin{multline}
    \text{Res}_{\tilde{X}_{2,5}(z^{*})=0}  = \frac{(\tilde{X}_{2,4} + \tilde{X}_{3,5})(\tilde{X}_{1,5} + \tilde{X}_{2,6})}{\tilde{X}_{2,5}} \, + \\ (\tilde{X}_{2,5} - \tilde{X}_{2,4} - \tilde{X}_{3,5} - \tilde{X}_{1,5} - \tilde{X}_{2,6}).
\end{multline}
The contribution from the pole at $\tilde{X}_{3,6}(z) = 0$ similarly gives
\begin{multline}
   \text{Res}_{\tilde{X}_{3,6}(z^{*})=0}  = \frac{(X_{1,3} + X_{2,6})(X_{3,5} + X_{4,6})}{X_{3,6}} \, + \\ \hf (X_{1,5} - X_{1,4} - X_{4,6} + 2X_{2,4} - 2X_{2,5}).
\end{multline}
Summed together these give the correct six-point mNLSM amplitude in terms of planar variables:
\begin{multline}
    A_{6}^{\tx{mNLSM}} = \frac{(\tilde{X}_{1,3} + \tilde{X}_{2,4})(\tilde{X}_{1,5} + \tilde{X}_{4,6})}{\tilde{X}_{1,4}} \, + \\  \frac{(\tilde{X}_{2,4} + \tilde{X}_{3,5})(\tilde{X}_{1,5} + \tilde{X}_{2,6})}{\tilde{X}_{2,5}} \, + \\ \frac{(\tilde{X}_{1,3} + \tilde{X}_{2,6})(\tilde{X}_{3,5} + \tilde{X}_{4,6})}{\tilde{X}_{3,6}} \, - \\ \tilde{X}_{1,3} - \tilde{X}_{2,4} - \tilde{X}_{1,5} - \tilde{X}_{2,6} - \tilde{X}_{3,5} - \tilde{X}_{4,6} \, .
\end{multline}

\bibliography{biblio}

\end{document}